\documentclass[12pt]{article}
\usepackage{epsfig}
\usepackage{cite}
\begin{document}

\begin{titlepage}
\nopagebreak
{\flushright{
        \begin{minipage}{5cm}
        CERN-PH-TH/2008-019
        \end{minipage}        }
        
}
\vfill
\begin{center}
{ { \bf \sc \Large
Understanding the Standard Model, \\[0.2cm] as a bridge to the
discovery \\[0.2cm]
  of new phenomena at the LHC\footnote{Contribution to ``Perspectives
    on the LHC'', G. Kane and A. Pierce, editors.} }
}
\vfill                                                       
{\bf Michelangelo L. MANGANO$^{(a)}$ }
\\[0.5cm]
{\small
$^{(a)}$ CERN, Theoretical Physics Unit, \\
Physics Department, CH~1211 Geneva 23, Switzerland
}
\end{center}     
\nopagebreak
\vfill
\begin{abstract} 
I discuss the basic elements of the process that will lead to
the discovery of possible new phenomena at the LHC. We review the
status of the tools available to model the Standard Model backgrounds,
and the role that such tools can play in the discovery phase, and in
the exploration of the features and parameters of such new phenomena.
\end{abstract}                                                
\vskip 1cm
        CERN-PH-TH/2008-019 \hfill \\
\today \hfill  
\vfill       
\end{titlepage}

\def    \be             {\begin{equation}}
\def    \ee             {\end{equation}}
\def    \ba             {\begin{eqnarray}}
\def    \ea             {\end{eqnarray}}
\def    \nn             {\nonumber}
\def    \=              {\;=\;}
\def    \frac           #1#2{{#1 \over #2}}
\def    \ret            {\\[\eqskip]}
\def    \ie             {{\em i.e.\/} }
\def    \eg             {{\em e.g.\/} }
\def \lsim{\mathrel{\vcenter
     {\hbox{$<$}\nointerlineskip\hbox{$\sim$}}}}
\def \gsim{\mathrel{\vcenter
     {\hbox{$>$}\nointerlineskip\hbox{$\sim$}}}}
\def    \ev            {\mbox{$\mathrm{eV}$}}
\def    \kev            {\mbox{$\mathrm{keV}$}}
\def    \mev            {\mbox{$\mathrm{MeV}$}}
\def    \gev            {\mbox{$\mathrm{GeV}$}}
\def	\tev		{\mbox{$\mathrm{TeV}$}}
\def    \pt             {\mbox{$p_{\mathrm T}$}}
\def    \et             {\mbox{$E_{\mathrm T}$}}
\def    \et             {\mbox{$E_{\mathrm T}$}}
\def    \met            {\mbox{$\rlap{\kern.2em/}E_T$}}
\def    \ifb            {\mbox{$\mathrm{fb}^{-1}$}}
\def    \as             {\ifmmode \alpha_s \else $\alpha_s$ \fi}
\def    \asq             {\ifmmode \alpha_s(Q) \else $\alpha_s(Q)$ \fi}

\section{Introduction}
The Standard Model (SM) of fundamental interactions has by now been
successfully tested over the past 30 years, validating its dynamics
both in the gauge sector, and in the flavour structure, including a
compelling confirmation of the source of the observed violation of
parity (P) and combined charge and parity (CP) symmetries. The
inability of the SM to account for established features of our
universe, such as the presence of dark matter, the baryon asymmetry,
and neutrino masses, are not considered as flaws of the SM, but as
limitations of it, to be overcome by adding new elements, such as new
interactions and new fundamental particles.  With this perspective,
the LHC is not expected to further test the SM, but to probe, and
hopefully provide evidence for, the existence of such new
phenomena. Our ability to predict what will be observed at the LHC is
therefore not limited by fundamental issues related to left-over
uncertainties about the SM dynamics, but by the difficulty of
mastering the complex strong-interaction dynamics that underlies the
description of the final states in proton-proton collisions.

Many years of experience at the Tevatron collider, at HERA, and at
LEP, have led to an immense improvement of our understanding of this
dynamics, and put us today in a solid position to reliably anticipate
in quantitative terms the features of LHC final states. LEP, in
addition to testing with great accuracy the electroweak interaction
sector, has verified at the percent level the predictions of
perturbative QCD, from the running of the strong coupling constant, to
the description of the perturbative evolution of single quarks and
gluons, down to the non-perturbative boundary where strong
interactions take over and cause the confinement of partons into
hadrons. The description of this transition, relying on the
factorization theorem that allows to consistently separate the
perturbative and non-perturbative phases, has been validated by the
comparison with LEP data, allowing the phenomenological parameters
introduced to model hadronization to be determined. The factorization
theorem supports the use of these parameters for the description of
the hadronization transition in other experimental environments. HERA
has made it possible to probe with great accuracy the short-distance
properties of the proton, with the measurement of its partonic content
over a broad range of momentum fractions $x$. These inputs, from LEP
and from HERA, beautifully merge into the tools that have been
developed to describe proton-antiproton collisions at the Tevatron,
where the agreement between theoretical predictions and data confirms
that the key assumptions of the overall approach are robust. Basic
quantities such as the production cross section of $W$ and $Z$ bosons,
of jets up to the highest energies, and of top quarks, are predicted
theoretically with an accuracy consistent with the known experimental
and theoretical systematic uncertainties. This agreement was often
reached after several iterations, in which both the data and the
theory required improvements and reconsideration.  See, for example,
the long saga of the bottom-quark cross section~\cite{Mangano:2004xr}\,,
or the almost embarassing --- for theorists --- case of the production
of high transverse momentum $J/\psi$s~\cite{Abe:1992ww}\,.  

While the
present status encourages us to feel confident about our ability to
extrapolate to the LHC, the sometimes tortuous path that led to this
success demands caution in assuming by default that we know all that
is needed to accurately predict the properties of LHC final
states. Furthermore, the huge event rates that will be possible at the
LHC, offering greater sensitivity to small deviations, put stronger
demands on the precision of the theoretical tools. In this essay I
discuss the implications of these considerations, in the light of some
lessons from history, and I discuss the role that theoretical
calculations should have in the process of discovering new physics.  I
shall not provide a systematic discussion of the state of the art in
calculations and Monte Carlo tools (for these, see
Refs.~\cite{Dixon:2007hh} and~\cite{Dobbs:2004qw}, respectively) but
rather a personal perspective on aspects of the relation between
theory and data analysis that are sometimes neglected.  Furthermore, I
shall only deal with what we would call ``direct discovery'', namely
the observation of the production of some new particle. Of course the
LHC can discover new physics in other ways, for example by measuring
$B_s\to \mu^+\mu^-$ decays with a rate different than predicted by the
SM, or by measuring the top, $W$ and Higgs masses to be inconsistent
with the SM expectation. I shall not cover these aspects, since they
are subject to sources of theoretical and experimental uncertainties
that are rather complementary to those I intend to focus on.

\section{Signals of discovery}
Three new elementary particles have been discovered by hadron
colliders: the $W$ and $Z$
bosons~\cite{Arnison:rp,Banner:1983jy}\,, and the top
quark~\cite{Abe:1994st,Abachi:1995iq}\,. For the first two, the
features of the final states were known in advance with great
confidence, and so were the masses and the production rates. The
signals stood out of the backgrounds very sharply and cleanly,
and their interpretation in terms of $W$ and $Z$ was
straightforward. The discovery of the top was harder, but still
benefited from the a-priori knowledge that the top {\it had} to be
there somewhere, and of its production and decay properties.

It is likely that the search and discovery of the Higgs boson will
follow a similar path. We have reasonable confidence that the Higgs
{\it has} to be there, and we know how it would be produced and decayed, as a
function of its mass and even as a function of the possible models
alternative to the plain SM implementation of the Higgs
mechanism. Search strategies have been set up to cover all expected
alternatives, and in many cases a signal will be unmistakable: mass
peaks such as those obtained from $H\to \gamma \gamma$, or $H\to ZZ\to
4$ leptons, are easily established as soon as the statistics is large enough
to have them stand out of the continuum background, without any need
to rely on theoretical modeling. 

As we move away from the default Higgs scenarios, into the territory
of new physics beyond the SM, life becomes more difficult.  One should
think of two phases for a discovery: establishing the deviation from
the SM, and understanding what this deviation corresponds to. It is
crucial to maintain these two phases separate. The fact that a given
anomaly is consistent with one possible interpretation does not
increase its significance as an indication of new physics. If we see
something odd in a given final state, it is not by appealing to, or
freshly concocting, a new physics model that gives rise to precisely
this anomaly that makes the signal more likely or more credible.  The
process of discovery, namely the detection of a deviation from the SM
by more than, say, 5 standard deviations of combined statistical and
systematic uncertainties, should be based solely on the careful
examination of whether indeed this signal violates the SM
expectation. Assigning this discrepancy to a slot in the space of
possible BSM scenarios is a subsequent step.

We can broadly group the possible deviations from the SM expectation
into three, possibly overlapping, classes: mass peaks, shape
discrepancies, and excesses in so-called counting experiments.

\subsection{Mass peaks}
Whether in a dilepton, diphoton, or dijet final state, a two-body mass
peak in the region of hundred GeV and above is the most robust
signature one can hope for.  Unless one sculpts the signal with a
dangerous choice of selection cuts (like looking for a mass peak in
the mass region just above the kinematical threshold set by twice the
minimum energy of the reconstructed objects), this signal cannot be
faked by a detector flaw. For example, things like malfunctioning
calorimeter units occasionally giving a fixed signal corresponding to
a high-energy deposition, will only fake a mass peak if all events have
precisely the same two detector elements giving the signals for the
two particles in the mass bin. Random failures by more calorimeter
towers would give different two-body invariant masses, because of the
different reconstructed kinematics, and would not build up a mass
peak! On the other hand if it is always the same two calorimeter
towers giving the signal, this is unlikely enough to be immediately
spotted as a localized hardware problem rather than as a $Z'$.

On the theory side, no SM background can give rise to a sharp peak,
since, unless you are sitting on top of a $W$ or $Z$, 
 all sources give rise to a continuum spectrum: either from the
obvious DY, or from the  decay of separate objects (like $WW$ or
 $t\bar{t}$ pairs).

An experimental analysis would extract the background directly from
the data, by studying the sidebands of the invariant mass distribution
below and above the peak, and interpolating under it. The role of the
simulation of the SM background is therefore marginal, and will only
contribute, possibly, in helping the interpolation and establishing
more accurately the background level for the experimental extraction
of the signal excess. The simulation becomes then crucial in the
second phase, that of the determination of the origin of the new
signal, and of the study of its properties, as discussed later. 

Mass peaks are just an example of a general set of
self-calibrating signals, for which data themselves offer the most reliable
source of background estimate. Other examples include jacobian peaks,
or sharp edges in two-body mass distributions, like in the case of
dileptons in supersymmetric chain decays of gauginos~\cite{Hinchliffe:1996iu}\,.

\subsection{Anomalous shapes of kinematical distributions}
Typical examples in this category are the inclusive transverse energy
(\et) spectrum of jets, or the missing transverse energy (\met)
distribution in some class of final states (e.g. multijet plus \met,
as expected in most supersymmetric scenarios). A precise knowledge of
the SM background shapes is an obvious advantage in these cases. To
which extent one can solely rely on such presumed knowledge, however,
is a matter worth discussing. To help the discussion it is useful to
consider a concrete example from the recent history of hadronic
collisions, namely the high-\et\ jet spectrum measured in run~1 by
CDF~\cite{Abe:1996wy}\,.
 
\begin{figure}
\centerline{\psfig{file=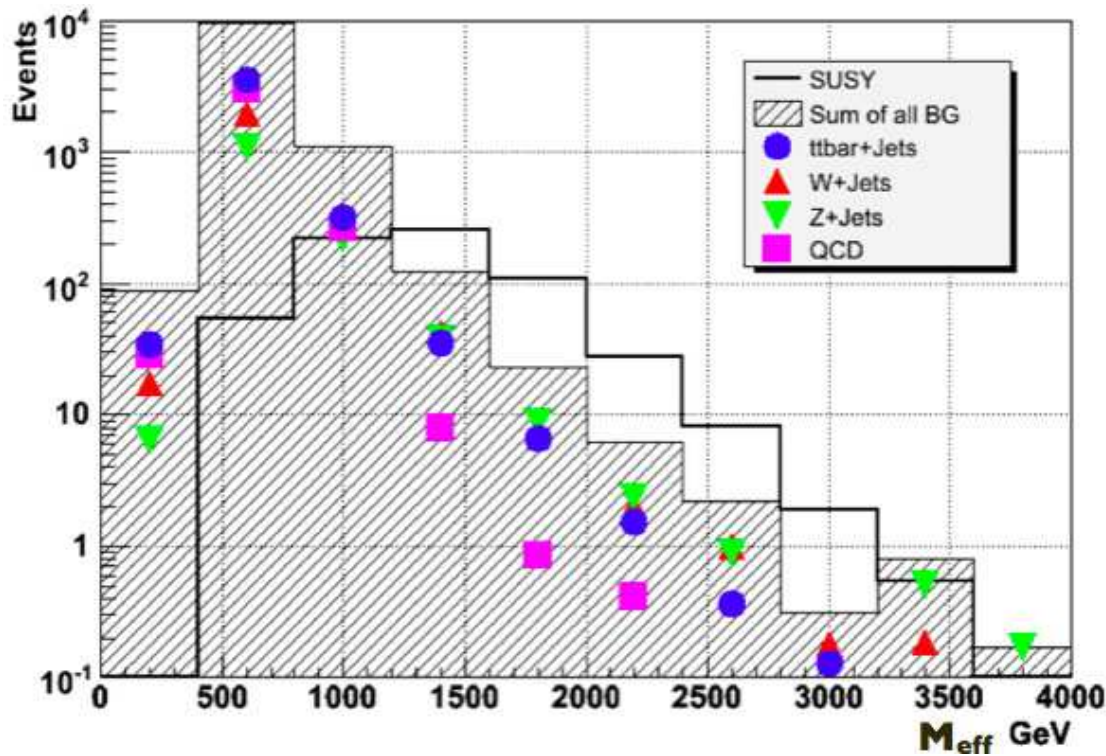,width=0.9\textwidth}}
\caption{Example of an expected supersymmetry signal and backgrounds in the
  multijet+\met\ final state~\cite{Vahsen}\,.}
\label{fig:met}
\end{figure}
For years it has been claimed that a high tail in the jet \et\
spectrum is a possible signal of an anomalous quark form factor, due
for example to the manifestation of quark compositeness. The
comparison of CDF's data with the best available theoretical
predictions, based on next-to-leading-order (NLO) QCD and the latest
parton density functions (PDFs) extracted from HERA, led to a
several-$\sigma$ excess in the region $\et\gsim 250$~GeV, compatible
with compositeness at a scale just above the TeV. The difference
between theory and data was such that no appeal to yet higher order
effects could have possibly fixed it. In that respect, the discrepancy
would have already been visible using a plain leading-order (LO)
calculation, since, aside from an overall $K$ factor, the shapes at LO
and NLO were known to agree very well. On the side of PDFs, the same
conclusion could be drawn from the analysis of the HERA data. With the
lack of flexibility in modifying the behaviour of the partonic NLO
cross section, and given the solid understanding of experimental
systematics, the PDFs remained however the only possible scapegoat. As
it turned out, the supposedly well-known large-$x$ behaviour of the
gluon density was driven mostly by the assumed functional form used in
the fits, rather than by data directly sensitive to it. Including the
CDF data into the fits, led in fact to new
parameterizations~~\cite{Lai:1996mg} that gave equally good
descriptions of previous data, as well as explaining away the jet
anomaly. The impasse was finally resolved by the subsequent analysis
by D0~\cite{Abbott:2000ew}\,, which considered the \et\ spectrum of
jets produced at large rapidity. The production of dijet pairs with a
large longitudinal boost but of low invariant mass (therefore in a
region free of possible new physics {\it contamination}) forces the
momentum fraction of one of the incoming partons to be close to 1,
thus probing the PDFs in the range relevant for the high-\et, central,
production. This measurement confirmed the newly proposed fits, and
set the matter to rest.

The lesson for the future is that, more than accurate theoretical
calculations, in these cases one primarily needs a strategy for an internal
validation of the background estimate. If evidence for some new phenomenon
entirely depends on the shape of some distribution, however accurate
we think our theoretical inputs are, the conclusion that there is new
physics is so important that people will always correctly argue
that perhaps there is something weird going on on the theory side, and
more compelling evidence has to be given. 

A place where we shall (hopefully!)  encounter this problem at the LHC
is the \met\ spectrum in multijet events, the classic signature of
escaping neutralinos produced in the chain decay of pair-produced
squarks and gluinos.  A possible outcome of this measurement is given
in Fig.~\ref{fig:met}, taken from recent ATLAS
studies~\cite{Vahsen}\,. The variable $M_\mathrm{eff}$ is defined as the
sum of the transverse energies of all hard objects in the event (jets
and \met, in this case). Events are required, among other things, to
have at least 4 jets with $\et>50\gev$, of which one above 100~GeV,
and $\met>\rm{max}(100\gev,0.2\times M_\mathrm{eff})$.  The solid
histogram is the expected signal, the shaded one is the sum of all
backgrounds, including SM processes with real \met (such as jets
produced in association with a $Z$ boson decaying to neutrinos), and
SM processes where the \met\ results from the inaccurate measurement
of some jet energies. The signal corresponds to production of squarks
and gluinos with a mass of the order of 1~TeV.
While the signal has certainly a statistical significance sufficient
to claim a deviation from the SM, it is unsettling that its shape is
so similar to that of the sum of the backgrounds. The theoretical
estimates of these backgrounds have also increased significantly over the
last few years, as a result of more accurate tools to describe
multijet final states. There is no question, therefore, that unless
each of the background components can be separately tested and
validated, it will not be possible to draw conclusions from the mere
comparison of data against the theory predictions. 

I am not saying this because I do not believe in the goodness of our
predictions. But because claiming that supersymmetry exists is far too
important a conclusion to make it follow from the straight comparison
against a Monte Carlo. One should not forget relevant examples from
the colliders' history~\cite{Arnison:1984iw,Arnison:1984qu}\,, such as
the misinterpretation in terms of top or supersymmetry of final states
recorded by UA1 with jets, \met, and, in the case of top,
leptons. Such complex final states were new experimental
manifestations of higher-order QCD processes, a field of phenomenology
that was just starting being explored quantitatively. It goes to the
theorists' credit to have at the time played devil's
advocate~\cite{Ellis:1985ig}\,, and to have improved the SM predictions,
to the point of proving that those signals were nothing but bread and
butter $W$ or $Z$ plus multijet production. But the fact remains that
claiming discoveries on the basis of a comparison against a MC is dangerous.

So let me briefly discuss the current status of theory
predictions for the SM channels relevant for supersymmetry
searches. There are three dominant processes: production of jets and a
$Z$ boson, with $Z\to \nu\bar\nu$ giving the missing energy;
production of jets and a $W$ boson, where this decays either to a
$\tau\nu$ (the $\tau$ faking a jet), or to a $\mu\nu$ or $e\nu$, with
the leptons escaping identification; and $t\bar{t}$ pairs, where one
of the $W$s from the top decays behaves like in the previous case.

$t\bar{t}$ production has been well tested at the
Tevatron~\cite{Sorin:2007zz,Yao:2006px}\,. 
Theoretical NLO calculations, enhanced
by the resummation of leading and subleading Sudakov
logarithms~\cite{Bonciani:1998vc}\,, predict correctly the total cross
section. The predictions for the LHC are expected to be equally
accurate, if not more, since the main source of uncertainty, the PDFs,
fall at the LHC in a range of $x$ values where they are known with
precision better than at the Tevatron. The kinematical production
properties, such as the transverse momentum distribution or the invariant
mass of the $t\bar{t}$ pair, are also well described by theory, and
Monte Carlo event generators are available to model the full structure
of the final states, including both the full set of NLO
corrections~\cite{Frixione:2007nw} and the emission of multiple extra
jets~\cite{Mangano:2006rw}\,, which is relevant for the backgrounds to
supersymmetry.
\begin{figure}
\centerline{\psfig{file=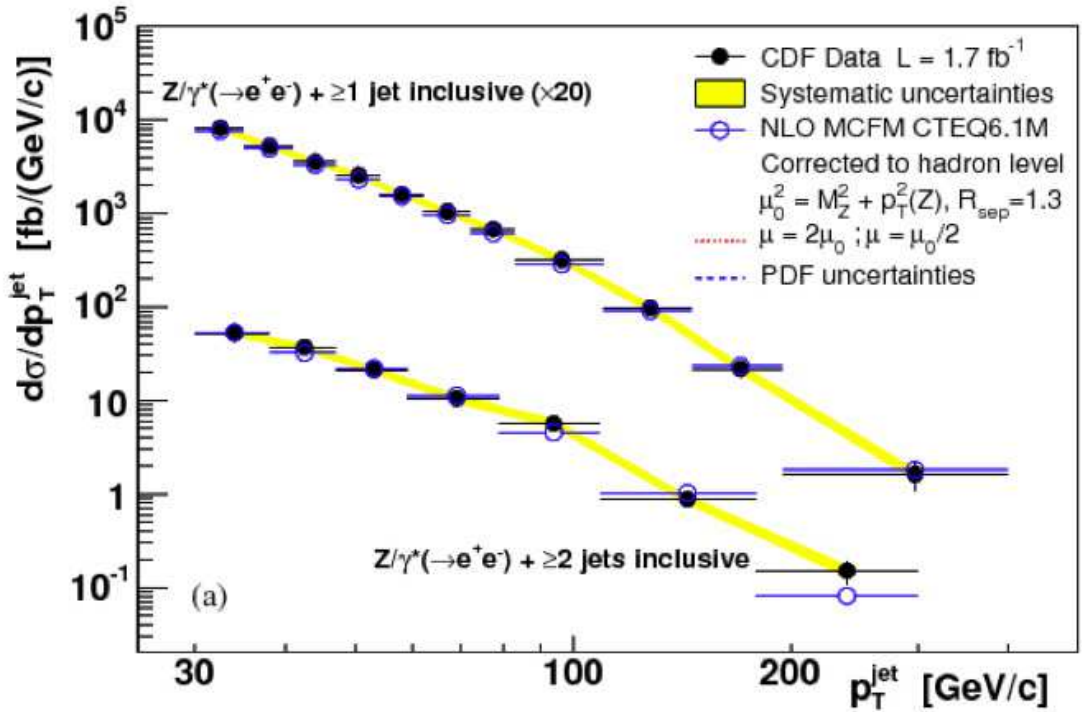,width=0.9\textwidth}}
\caption{Jet \et\ spectra in $Z$+jet(s), as measured by CDF at the
  Tevatron~\cite{Aaltonen:2007cp}\,.} 
\label{fig:zjet}
\end{figure}

The processes $W$+jets and $Z$+jets are very similar from the point of
view of QCD. There are minor differences related to the possibly
different initial-state flavour compositions, but the main theoretical
systematics, coming from the renormalization-scale sensitivity due to
the lack of higher-order perturbative corrections, are strongly
correlated. In the case of $W/Z$+1 and 2 jets, parton-level NLO
calculations are available~\cite{Campbell:2002tg}\,. They are in
excellent agreement with the measurements at the
Tevatron~\cite{Aaltonen:2007cp,Abazov:2006gs}\,, as shown
for example in the case of $Z$+1 and 2 jets by the CDF
results~\cite{Aaltonen:2007cp} shown
in Fig.~\ref{fig:zjet}.
Going to higher jet multiplicities, and generating a realistic
representation of the fully hadronic final state, is then possible
with LO calculations. Exact, LO matrix-element calculations of
multiparton production can be enhanced by merging with
shower Monte Carlo codes~\cite{Corcella:2000bw,Sjostrand:2003wg}\,,
which add the full perturbative gluon shower and
eventual hadronization. An example of the quality of these predictions
is given by Fig.~\ref{fig:wjet},  which shows the ratio of the
measured~\cite{Aaltonen:2007ip} and predicted $W+N$-jet cross
sections, for jets with $\et>25$~\gev.
The theoretical predictions include the LO results from
Ref.~\cite{Mangano:2002ea} (labeled as MLM), and from
Ref.~\cite{Mrenna:2003if} (labeled as SMPR), while MCFM refers to the
NLO predictions for the 1- and 2-jet rates from
Ref.~\cite{Campbell:2002tg}\,.  The systematic uncertainties of the
individual calculations, mostly due to the choice of renormalization
scale, are shown.  The LO results, which have an absolute
normalization for all $N$-jet values, are in good agreement
with the data, up to an overall $K$ factor, of order 1.4. The
prediction for the ratios of the $N$-jet and $(N-1)$-jet rates is also
in good agreement with the data. The NLO calculations embody the $K$
factor, and exactly reproduce the 1- and 2-jet rates.
\begin{figure}
\centerline{\psfig{file=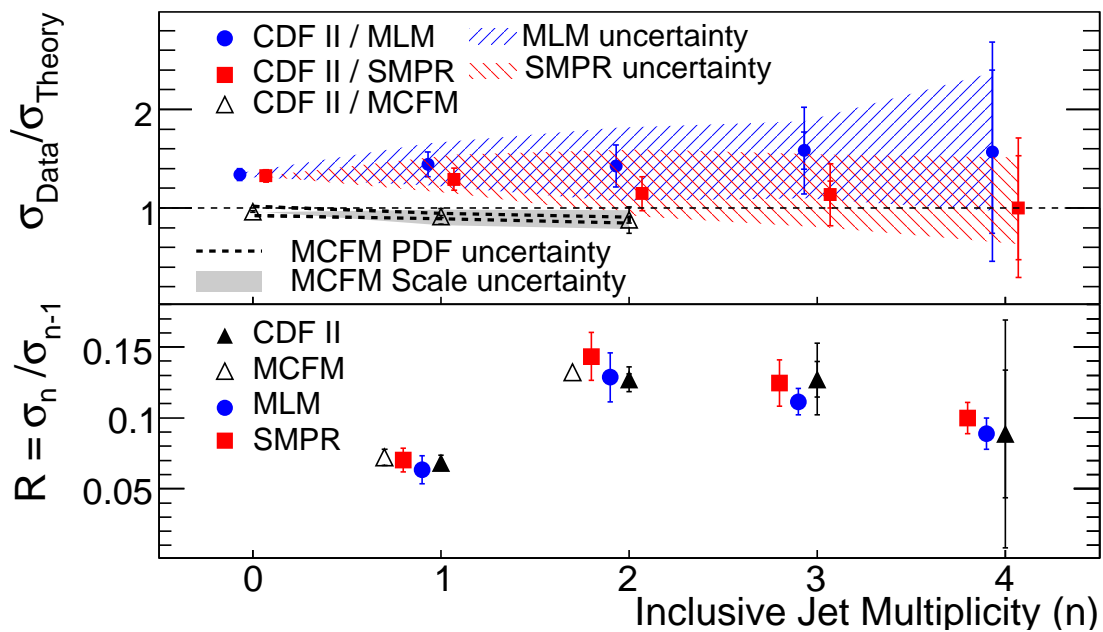,width=0.9\textwidth}}
\caption{Comparison between CDF data and theory for $W+N$ jets cross
  sections~\cite{Aaltonen:2007ip}\,.}
\label{fig:wjet}
\end{figure}

Thorough comparisons have been performed~\cite{Alwall:2007fs} among a
set of independent calculations of $W$ plus multijet final
states~\cite{Krauss:2004bs,Lavesson:2005xu,Maltoni:2002qb,
Papadopoulos:2005ky,Mangano:2002ea}\,. 
The results of the matrix element evaluation
for these complex processes are all in excellent agreement;
differences in the predictions at the level of hadrons may instead
arise from the use of different parton-shower approaches, and of
different ways of sharing between matrix elements and shower the task
of describing the radiation of hard jets. An example of the spread in
the predictions is shown in Fig.~\ref{fig:et}, which shows the \et\
spectra of the four highest-\et\ jets in $W$+multijet events at the
LHC. With the exception of the predictions from one of the codes, all
results are within $\pm 50\%$ of each other, an accuracy sufficient by
itself to establish possible deviations such as those in
Fig.~\ref{fig:met}.
\begin{figure}
\centerline{\psfig{file=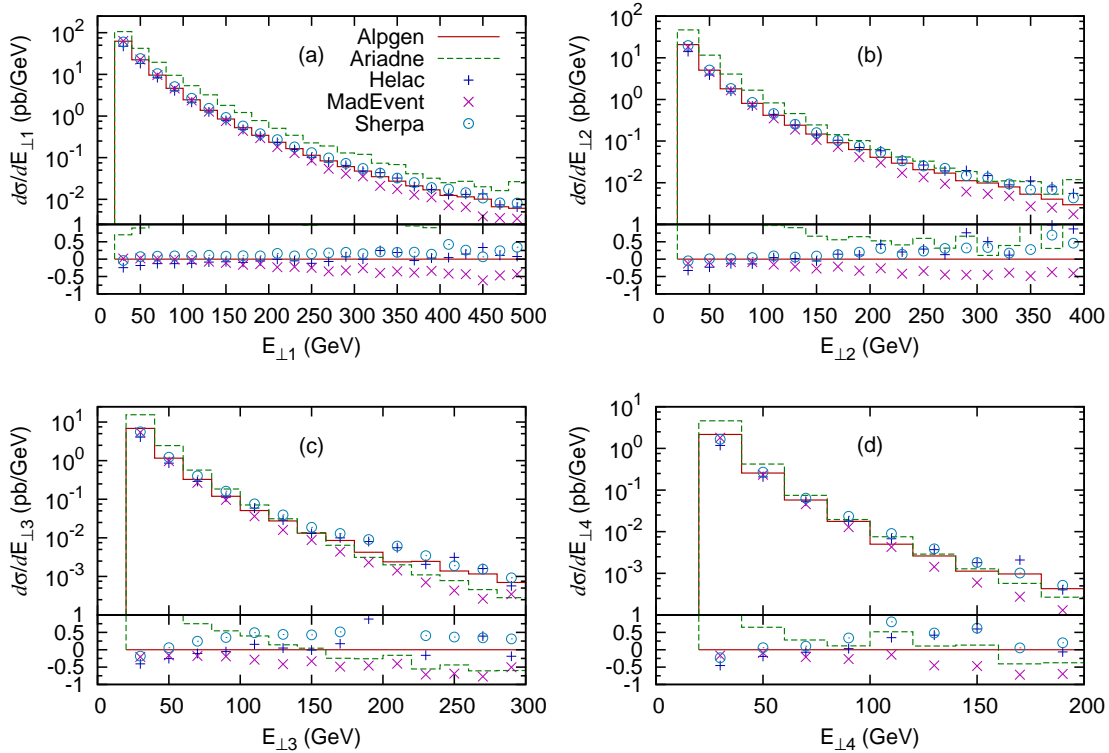,width=0.9\textwidth}}
\caption{Predicted jet \et\ spectra in $W$+jet(s) final states at the
  LHC~\cite{Alwall:2007fs}\,.} 
\label{fig:et}
\end{figure}
These differences are of size compatible with the intrinsic
uncertainties of the calculations, given for example by the size of
the bands in Fig.~\ref{fig:wjet}. It is expected that they can be
removed by tuning the input parameters, like the choice of
renormalization scale, by fitting the data. An accurate determination
of the normalization and shape of the SM background to a
supersymmetric signal could therefore be obtained by analyzing data
control samples. The description of the $(Z\to\nu\bar{\nu})+$ jets
process can be validated, and the absolute normalization of the rate tuned, 
by measuring the signal-free
$(Z\to e^+e^-)+$ jets final states. This information can be then
directly used to tune the $W$+jets predictions; or one can measure
directly $(W\to e \nu)+$ jets in a region where the electron is
clearly tagged, and use the resulting tune to extrapolate to the case
of $\tau$ decays, or to decays where the $e$ and $\mu$ are not
detected. The fact that the calculations appear to well reproduce the
ratios of $\sigma[N\mathrm{-jet}]/\sigma[(N-1)\mathrm{-jet}]$,
provides a further handle. 

A clear path is therefore available to establish the accuracy of the
theoretical tools, and to provide robust background estimates for
searches of anomalies in the multijet plus \met\ final states. As
always, however, the devil is in the details. As shown by the Tevatron
analyses, even the measurement of the background $W$+multijet cross
sections is not an easy task, due to a large contamination from
$b\bar{b}$ backgrounds (where both $b$-hadrons decay
semileptonically, one giving rise to a hard and isolated charged
lepton, the other to a very energetic neutrino), and $t\bar{t}$
backgrounds, which at the LHC are the dominant source of $W$+multijet
events. It is therefore difficult to anticipate the dimension of the
challenge, only the direct contact with to data will tell! 

\subsection{Counting experiments}
A {\it counting experiment} is a measurement defined by assigning some
selection criteria, counting the number of events passing the cuts,
and comparing this against the expected background.  Counting
experiments are like searches for shape anomalies, since the analysis
cuts act on the distribution of some variables. However the
expected statistics are too small to allow a meaningful use of the
full distributions, and one simply integrates over the full sample
passing the cuts. So counting experiments tend to lack a {\it smoking
  gun}, a truly compelling evidence that something is wrong, and they
require the most robust understanding of backgrounds one can possibly
need.

A typical example is given by the analyses that led
to the top discovery. Different selections were applied to single out
complementary data samples, each characterized by at least one of the
expected features of top final states: 
\begin{itemize}
\item a charged lepton, with \met, 3 or more jets, and possibly one
  of them containing either a lepton, or a secondary displaced vertex (SVX),
  expected features of $b$-hadron decays;
\item a pair of opposite-charge leptons, with invariant mass away
  from the $Z$ peak is same flavour, and one or two jets, possibly
  with a $b$-quark tag.
\end{itemize}
Each of the objects listed above had to pass some kinematical or
quality cuts, in
terms of minimum \pt\ or \et, or in terms of variables defining the
cleanness of the leptonic of SVX tags. The
estimate of the backgrounds to a counting experiment is usually very
hard. One can always suspect that, even if the background estimates are
tested on control samples, the extrapolation in the signal region
could fail. For example, backgrounds that are a negligible component
of the control samples could sneak into the signal region, and
suddenly dominate the rates. Furthermore, backgrounds could have
strong correlations among different variables, and the probability of
extrapolating their rate using their relative weight for various
variables may not factorize into a product of probabilities. A typical
example is the one given earlier for the contribution of $b\bar{b}$
pairs to events with isolated electrons (or muons) and missing
transverse energy. The contribution to isolated leptons is
proportional to the probability of having a $b\bar{b}$ final state,
$P_{bb}$, times the probability of a suitable decay to a charged
lepton, $P_{b\to\ell}$. The contribution to a large \met\ signal is
likewise given by $P_{bb}\times P_{b\to\nu}$. But the probability of
having both a hard charged lepton and the \met\ is not the product of
the two, but is given by $P_{bb}\times P_{b\to\ell} \times
P_{b\to\nu}$, which is larger than the product by $\sim 1/P_{bb} \gsim
100$.

Analyses of this type require full Monte Carlo codes, where the best
possible perturbative input (e.g. multijet matrix elements) is used
together with a complete description of the shower, hadronization,
particle decays, and underlying event. A good description of a
$b$-hadron decay, based for example on an empirical fit to
existing data, may have more value than the inclusion of
first-principle NLO 
corrections to the matrix elements. The challenge for the theorist is
to provide a prediction where the accuracy is uniformly distributed over all
the delicate areas, rather than concentrated on some specific
spots. The calculation must have enough tunable parameters that the
predictions can be adjusted to fit the data in the control samples, but
not too many that the extrapolation from the control samples to the
signal regions may not be trustable. The crucial question that a
theorist is called to answer for applications to counting experiments
is, in fact, to which extent the predictions of the code tuned on some
sample can be trusted when exported to another sample. When codes can
be properly tuned, the {\it portability} issue becomes the dominant
source of systematics. Sorting out all of these issues requires a very
careful and skilled work, both on the experimental and theoretical
side. This sort of explains why it took about 100 pages~\cite{Abe:1994st} to
document the steps that led to a credible first evidence for the top quark!

The discovery of the Higgs in complex final states, such as the
weak-boson fusion channels, with the Higgs decaying to final states
without a sharp mass peak, such as $H\to \tau^+\tau^-$ or $H\to
b\bar{b}$, and with vetoes on the presence of jets in the central
rapidity region, will fall in this category of extremely difficult and
hard-to-valiate searches.

An interesting historical example in this class is the famous
$e^+e^-\gamma\gamma \met$ event seen by CDF in
run~1~\cite{Abe:1998ui}\,.  The expected SM background for this event is
less than $10^{-6}$ events. The estimate drops to $10^{-8}$ if one
assumes that the most forward electron (for which the tracking
information, and therefore a charge-tag, is missing) is actually a
photon. According to {\it the rules}, this is a 5$\sigma$ excess, and
qualifies for a discovery.  After all, even the $\Omega^-$ discovery
was based on just one, compelling, event. In terms of pure statistics,
the $e^+e^-\gamma\gamma \met$ is (still today, after 30 times more
luminosity has been collected by CDF and D0) even more significant as
a deviation from the background (whether caused by physics or
instrumental) than the first $W$ observations at UA1 and UA2.  Why do
we not consider it as evidence of new physics?  Because consensus
built up in the community that, in spite of the ``5$\sigma$'', the
evidence is not so compelling. On one side plausible BSM
interpretations have been ruled out~\cite{Yao:2006px}
 by the LEP experiments, which
inconclusively explored, for example, scenarios based on
gauge-mediated supersymmetry breaking, with $\tilde{e} \to
e\tilde{\gamma} \to e \gamma \tilde{G}_{3/2}$. On the other, doubts
will always remain that some freaky and irreproducible detector effect
may be at the origin of this event. However difficult, the estimate of
the physics SM background to this event at the leading-order is
relatively straightforward and has been checked and
validated. Higher-order effects, not known, cannot be reasonably
expected to change the rates by more than a factor of two. I do not
think that anyone can seriously argue that the knowledge of the
background rates with a NLO accuracy would have changed our
conclusions about this event, so I do not think that here theory could have
played a more important role. As in other examples of the past (most
frequently in the discovery of hadronic resonances, see e.g. the
recent case of pentaquarks), theoretical bias (e.g. the availability
and appeal of a theoretical framework  --- or lack thereof --- within
which to fit the claimed discovery), the possible prejudice towards
the robustness of the analysis or of the group that performed it, and
other considerations generically labeled as {\it guts' feeling},
heavily interfere with the purely statistical and systematics
assessment of a finding, making its interpretation more difficult. We
find a similar situation in other areas of particle physics. The
examples of neutrino oscillations and of the muon anomalous magnetic
moment come to mind. Davies' solar-neutrino anomaly had been sitting
there for years, and no improvement in the solar model was ever going to be
good enough for the reality of neutrino oscillations to be uniformly
accepted by the community. New data, less sensitive to the details of
theory, and providing the opportunity to test more convincingly the
model assumptions beyond the shade of any doubt, had to come for
Davis' signal to be incorporated in a broader evidence for neutrino
oscillations. It is likely that the 3.5$\sigma$ of BNL's $g_\mu-2$
experiment~\cite{Bennett:2006fi} will have a similar fate, regardless
of how much progress will be made in the theoretical understanding of
the hadronic contribution to light-by-light scattering. QCD is simply
too vicious for everyone to accept that this anomaly is a conclusive
evidence of physics beyond the SM, let alone to commit to an
interpretation such as supersymmetry. 
 
\section{Measuring parameters}
A key element of the discovery programme at the LHC will be improving
the accuracy of the SM parameters, and measuring, as precisely as
possible, the parameters of the new physics that will hopefully be
discovered. The relation between $m_{top}$, $m_W$, $\sin\theta_W$ and
$M_H$ is an important prediction of the SM, and deviations from it
should be accounted for by the effects of new physics. And in presence of
new physics, the values of the new particles' masses and couplings
will be the starting point to reconstruct the new theory.

This is an area where the ability of theory to describe the final
states is crucial. Couplings will be extracted from the determination
of production cross sections, branching ratios, or angular
distributions. Masses will mostly be obtained via direct kinematical
reconstructions. In all cases, an accurate modeling of both the SM
backgrounds, which contaminate and deform the signal distributions,
and of the signals, will be required.

Cross sections are obtained by counting events. Since the analyses
defining a given signal have always selection cuts, to go from event
counts to a cross section one has to model the acceptance and
efficiency of those cuts. These depend on the details of the
production process, something that only a theoretical calculation can
provide. This implies that the calculations should not only provide a
precise value of the total cross section of a given process, but also
of the kinematical distributions that are used in the experimental
analysis. For example, in the case of the $W$ or $Z$ cross section one
needs the precise form of the \pt\ and rapidity spectra of the decay
leptons~\cite{Frixione:2004us}\,. The problem with this is that
typical higher-order calculations are more easily done at the total
cross section level, to benefit from the full inclusivity of the final
state and more easily enforce the cancellation of the divergencies
that appear separately in the real and in the virtual corrections.  A
great amount of work has therefore been invested recently in
developing techniques capable of delivering the same perturbative
accuracy both at the level of total cross sections and at the level of
distributions (for a review of recent developments in higher-order
perturbative QCD calculations, see e.g. Ref.~\cite{Dixon:2007hh}).
For example, the full next-to-next-to-leading-order (NNLO) calculation
of the lepton distributions in $pp\to (W\to \ell\nu)+X$ was recently
completed, in Ref.~\cite{Melnikov:2006di}\,. Their conclusion is that
the inclusion of NNLO corrections is necessary to control the rates at
the level of few percent. This is required, for example, for the
extraction of the LHC absolute luminosity at a similar level of
precision.  Such an accurate knowledge of the luminosity is the
prerequisite for the precise measurement of all other cross sections,
including those of interesting new processes. More recently, even the
calculation of rates for some Higgs final states has reached a full
NNLO precision for realistic leptonic
observables~\cite{Anastasiou:2007mz}\,.

Purely leptonic observables, where the leptons arise from the decay of
non-strongly interacting particles, make it possible to fully
integrate over the strongly interacting components of the events and,
experimentally, enjoy a reduced dependence on the full hadronic
structure. Under these circumstances, the use of parton-level
calculations for realistic studies is legitimate (see
e.g. Ref.~\cite{Anastasiou:2008ik} for a discussion of $pp\to H\to WW\to
\ell\ell\nu\nu$).

Precision measurements of observables directly sensitive to the
hadronic component of the events are typically more demanding.  A good
example of the difficulties that are encountered in these cases is
given by the measurement of the top quark mass. In hadronic collisions
the top quark mass can only be measured by reconstructing, directly or
indirectly, the total invariant mass of its decay products. Due to the
large phase-space available, top quark pairs are always produced well
above their kinematical threshold. One cannot therefore use techniques
such as those available in $e^+e^-$ collisions, where the mass of a
new particle can be deduced from an energy scan at the production
threshold. Furthermore, contrary again to $e^+e^-$ collisions where a
top-quark pair at threshold is produced without any other object, in
the $pp$ case the top pair is always accompanied by both the fragments
of the colliding protons, and by the multitude of hadrons that are
radiated off as the incoming quarks or gluons that will fuse into
$t\bar{t}$ approach each other (initial-state radiation). It is
therefore impossible to exactly decide which ones among the many
particles floating around originate from the top decays, and have to
be included in the determination of the top invariant mass, and which
ones do not. As an additional obstacle, the top quark is coloured, it
decays before hadronizing, but the detected decay products must be
colour-singlet hadrons. This implies that at some stage during the
evolution of the quarks and gluons from the top decay they will have
to pair up with at least one antiquark drawn from the rest of the
event, to ensure the overall colour neutrality of the final
state. There is therefore no way, as a matter of principle, that we
can exactly measure on an event-by-event basis the top quark mass. The
best that we can do is to model the overall production and decay
processes, and to parameterize a set of determined observables as a
function of the input top quark mass. For example, such an observable
could be the invariant mass of three jets, assuming one of them comes
from the evolution of the $b$ quark, and the other two from the decay
of the $W$. This modeling cannot be achieved with parton-level tools,
regardless of their perturbative accuracy. A full description of the
final state is required, including the non-perturbative modeling of
both the fate of the proton fragments and of the transition turning
partons into colour-singlet hadrons. With the current level of
experimental uncertainties~\cite{:2007bxa} on $m_{top}$, at the 2 GeV
level, we are approaching the level where the theoretical
modeling~\cite{Skands:2007zg} is not validated by a direct comparison
with data. At the LHC, where the experimental uncertainties could be
reduced below the 1~GeV level, theory will be the dominant source of
systematics. Observables will have to be identified that will allow a
validation and tuning of this systematics, in the same way that
analogous problems had to be addressed for the determination of the
$W$ mass at LEP. This is an area where the Tevatron statistics are too
small to allow any progress, and all the work will be left to the
LHC. Needless to say, all of this work will benefit the precise
measurement of the masses of possible new particles decaying to
quarks and gluons.

\section{Conclusions}
Advanced MC tools for the description of the SM, and for the isolation
of possible new physics at the LHC, are becoming mature.  Validation
and tuning efforts are underway at the Tevatron, and show that a solid
level of understanding of even the most complex manifestations of the
SM are well under control.  The extrapolation of these tools to the
energy regime of the LHC is expected to be reliable, at least in the
domain of expected discoveries, where the energies of individual
objects (leptons, jets, missing energy) are of order 100~GeV and more. However,
the consequences of interpreting possible discrepancies as new physics
are too important for us to blindly rely on our faith in the goodness of the
available tools.  An extensive and coherent campaign of MC testing,
validation and tuning at the LHC will therefore be required.  Its
precise definition will probably happen only once the data are
available, and the first comparisons will give us an idea of how far
off we are and which areas require closer scrutiny.

Ultimately the burden, and the merit, of a discovery should and will
only rest on the experiments themselves! The data will provide the
theorists guidance for the improvement of the tools, and the analysis
strategies will define the sets of control samples that can be used to
prepare the appropriate and reliable use of the theoretical
predictions.

Aside from the discovery of anticipated objects like the $W$, $Z$ and
the top, we have never faced with high-energy colliders the concrete
situation of a discovery of something beyond the expected. In this
respect, we are approaching what the LHC has in store for us without a
true experience of discovering the yet unknown, and we should
therefore proceed with great caution.  All apparent instances of
deviations from the SM emerged so far in hadronic or leptonic
high-enegy collisions have eventually been sorted out, thanks to
intense tests, checks, and reevaluations of the experimental and
theoretical systematics.  This shows that the control mechanisms set
in place by the commonly established practice are very robust.

Occasionally, this conservative approach has delayed in some areas of
particle physics the acceptance of true discoveries, as in the case of
Davies's neutrino mixing, and as might turn out to be the case for the
muon anomaly. But it has never stopped the progress of the field, on
the contrary, it has encouraged new experimental approaches, and has
pushed theoretical physics to further improve its tools.

The interplay between excellent experimental tools, endowed with the
necessary redundancy required to cross-check odd findings between
different experiments and different observables, and a hard-working
theoretical community, closely interacting with the experiments to
improve the modeling of complex phenomena, have provided one of the
best examples in science of responsible and professional {\it modus
operandi}. In spite of all the difficult challenges that the LHC will
pose, there is no doubt in my mind that this articulated framework of
enquiry into the yet unknown mysteries of nature will continue
providing compelling and robust results.
\section{Acknowledgements}
I wish to thank C. Campagnari, A. Tollestrup and A. Yagil, who shared
with me over the years 
their wisdom on the topics touched by this note.

\end{document}